\documentclass[twocolumn,superscriptaddress,preprintnumbers,nofootinbib]{revtex4}
\usepackage{amsmath,amssymb,longtable}
\usepackage{graphicx}
\usepackage{epsfig,color}
\usepackage{bm}
\usepackage{verbatim}

\begin{document}

\title{$\beta$-decay rates of $^{(115,117)}$Rh into $^{(115,117)}$Pd isotopes in the microscopic IBFM-2}

\author{J. Ferretti}\email[]{jferrett@jyu.fi}
\affiliation{Department of Physics, University of Jyv\"askyl\"a, P.O. Box 35, 40014 Jyv\"askyl\"a, Finland}
\author{J. Kotila}\email[]{jenni.kotila@jyu.fi}
\affiliation{Finnish Institute for Educational Research, University of Jyv\"askyl\"a, P.O. Box 35, 40014 Jyv\"askyl\"a, Finland}
\address{Center for Theoretical Physics, Sloane Physics Laboratory, Yale University, New Haven, Connecticut 06520-8120, USA}
\author{R. I. Maga\~na Vsevolodovna}\email[]{rmagana@ge.infn.it}
\affiliation{INFN, Sezione di Genova, Via Dodecaneso 33, 16146 Genova, Italy}
\author{E. Santopinto}\email[]{elena.santopinto@ge.infn.it}
\affiliation{INFN, Sezione di Genova, Via Dodecaneso 33, 16146 Genova, Italy}

\begin{abstract}
The structure of odd-$A$ $^{(115,117)}$Rh and $^{(115,117)}$Pd isotopes is studied by means of the neutron-proton Interacting Boson-Fermion Model (IBFM-2). $J^P = \frac{1}{2}^+$ quantum number assignment for the $^{(115,117)}$Pd ground-states is critically discussed and the predicted energy levels are compared to the existing experimental data.
The resulting nuclear wave functions are used to compute the $\beta$-decay $ft$ values of the transitions from $^{(115,117)}$Rh to $^{(115,117)}$Pd in the microscopic IBFM-2 and the results compared with the data.
\end{abstract}

\maketitle

\section{Introduction}
$\beta$ and double-$\beta$ decays are manifestations of the weak interaction and are among the main mechanisms for the atomic nuclei to reach stability.
The investigation of such type of transitions, both on the theoretical and experimental sides, is extremely important. 
It can provide insight into the structure of heavy nuclei and serve as a test of the available nuclear structure models, including the Interacting Boson and Boson-Fermion Models (IBM-2 and IBFM-2) \cite{Navratil:1988zz,Dellagiacoma:1989fii,Yoshida:2002ju,Zuffi:2003iq,Brant:2004uj,Brant:2006iy,Barea:2013bz,Mardones:2016wgy,Nomura:2019qlb}, the Quasiparticle Random Phase Approximation (QRPA) \cite{Sarriguren:2001hb,Simkovic:2013qiy,Pirinen:2015sma,Marketin:2015gya,Mustonen:2015sfa,Nabi:2016saz}, and the large-scale nuclear shell model \cite{Langanke:2002ab,Caurier:2004gf,Yoshida:2018bog}. This is because the $\beta$-decay rates are very sensitive to the wave functions of the parent and daughter nuclei.

Furthermore, the study of $\beta$-decays can help to model the creation of chemical elements in the investigation of different possible astrophysical nucleosynthesis scenarios.
One can also mention the fundamental role of $\beta$, double-$\beta$ and neutrinoless double-$\beta$ decays in the study of the properties of neutrinos and of the possible emergence of beyond the Standard Model effects. Typical examples are the still unknown mechanism to generate a neutrino mass term in the Standard Model Lagrangian and the possible nature of neutrinos as Dirac- or Majorana-type particles \cite{Giunti:2003qt}.

A large amount of experimental data on $\beta$ decays has been collected during the years; for example, see Refs. \cite{Aysto:1988axd,Penttila-PhD,Dillmann:2003zz,Kurpeta:2010zza,Nishimura:2011zza,Rajabali:2012tq,Quinn:2011gs,Lorusso:2015cna,Caballero-Folch:2015iux,Kurpeta:2018zia}.
Here, the attention is focused on the experimental results for $^{115}$Rh and $^{117}$Rh isotopes decaying into $^{115}$Pd and $^{117}$Pd \cite{Aysto:1988axd,Penttila-PhD,Kurpeta:2010zza,Kurpeta:2018zia}.
The experimental study of these nuclei was motivated by the longstanding prediction of a shape change from prolate to oblate deformation expected in this region \cite{Skalski:1997cp,Xu:2002ry}, which is still a rare phenomenon in nuclei.
The first experimental signs of this change were reported in $^{110}$Mo \cite{Urban:2004} and $^{111}$Tc \cite{Urban:2005,Kurpeta:2011zz}. 
In the chain of Pd isotopes a hint of a transition to the oblate regime was reported in $^{115}$Pd \cite{Kurpeta:2010zza}. 
In the present paper, the energy levels of $^{115}$Rh, $^{117}$Rh, $^{115}$Pd and $^{117}$Pd nuclei and the Rh to Pd $\beta$-decay rates are studied in the microscopic IBFM-2. Results for the electromagnetic transitions are used to discuss the $J^P = \frac{1}{2}^+$ quantum number assignment for the $^{115}$Pd and $^{117}$Pd ground-states. 

The neutron-proton Interacting Boson-Fermion Model (IBFM-2) \cite{Iachello:2005aqa} is an extension of the well-known Interacting Boson Model (IBM-2) \cite{Iachello:2006fqa}.
The IBM-2 was originally introduced as a phenomenological approach to describe collective excitations in nuclei \cite{Arima:1976ky}.
Soon afterwards, however, its relation with the shell model was established \cite{Arima:1977vie,Otsuka:1978zza,Otsuka:1978zz}.
The IBM-2 deals with even-even nuclei, where one replaces valence nucleon pairs with bosons with angular momentum 0 or 2. By coupling an extra fermion to the previous boson system, one is able to extend the IBM-2 to the study of odd-$A$ nuclei. This extension of the model is known as IBFM-2.

In the IBFM-2 approach, $\beta$-decays are modeled as a combination of a neutron (proton) stripping and proton (neutron) pickup reactions \cite{Dellagiacoma:1989fii}. The previous process can be described in terms of a one-nucleon transfer operator, which is obtained following the method that avoids the use of Number Operator Approximation as discussed in
 \cite{Mardones:2016wgy,Matus:2017eni}, and then use it to calculate the $\beta$-decay rates of $^{(115,117)}$Rh  into $^{(115,117)}$Pd isotopes in the IBFM-2.

The paper is organized as follows: In Sec. \ref{sect2} we give the necessary theory background concerning IBFM-2 calculations and report our results for energy levels, whereas $M1$ transitions and ground state quantum number assignments are discussed in Sec. \ref{sect3} and beta decay rates in Sec. \ref{beta decays}. Finally, conclusions are drawn in Sec. \ref{conclusions}

\section{Spectrum of $^{(115,117)}$Rh and $^{(115,117)}$Pd isotopes in the IBFM-2}
\label{sect2}

\subsection{IBFM-2 Hamiltonian}
The Interacting Boson-Fermion Model (IBFM-2) \cite{Iachello:2005aqa} is an extension of the Interacting Boson Model (IBM-2) \cite{Iachello:2006fqa} to study even-odd nuclei.
Such odd-$A$ nuclei are described by coupling an odd nucleon (the fermion) to the even-even neutron-proton core (the bosonic system).

The IBFM-2 Hamiltonian is given by \cite{Iachello:2005aqa,Yoshida:2013jh}
\begin{equation}
	\label{eqn:H-IBFM}
	H = H^{\rm B} + H^{\rm F}_{\nu,\pi} + V^{\rm BF}_{\nu,\pi}  \mbox{ }.
\end{equation}	
Here, $H^{\rm B}$ is the IBM-2 Hamiltonian \cite{Otsuka:1978zz,Iachello:2006fqa,Kotila:2012zz}, which describes the even-even core nucleus, and for our purposes reads
\begin{widetext}
\begin{equation}	
	\label{eqn:H-IBM}
	\begin{array}{rcl}
	H_{\rm B} & = & \epsilon_d (\hat n_{d_\pi} + \hat n_{d_\nu}) + \kappa \left(Q_\nu^{\rm B}\cdot Q_\pi^{\rm B}\right)
	+ \frac{1}{2} \xi_2  \left[\left(d_\nu^\dag s_\pi^\dag - d_\pi^\dag s_\nu^\dag\right) \cdot 
	\left(\tilde d_\nu s_\pi - \tilde d_\pi s_\nu\right) \right] \\
	& + & \displaystyle \sum_{K=1,3} \xi_K \left[d_\nu^\dag \times d_\pi^\dag\right]^{(K)} \cdot 
	\left[\tilde d_\pi \times \tilde d_\nu\right]^{(K)} + \frac{1}{2} \displaystyle \sum_{K=0,2,4} c_\nu^{(K)} 
	\left[d_\nu^\dag \times d_\nu^\dag\right]^{(K)} \cdot \left[\tilde d_\nu \times \tilde d_\nu\right]^{(K)}
	\end{array} \mbox{ }.
\end{equation}
\end{widetext}
In the previous expression, $\hat n _{d_\rho} = d_\rho^\dag d_\rho$ and 
\begin{equation}
	\label{eqn:Q-rho}
	Q_\rho^{\rm B} = d_\rho^\dag s_\rho + s_\rho^\dag \tilde d_\rho + \chi_\rho [d_\rho^\dag \times\tilde d_\rho]^{(2)}
\end{equation}
represent the $d$-boson number operators and the boson quadrupole operators for the proton ($\rho = \pi$) and neutron ($\rho = \nu$) pairs, respectively; $s_\rho^\dag$ and $d_\rho^\dag$ are $s_\rho$- and $d_\rho$-boson creation operators, and the modified $d_\rho$-boson annihilation operator satisfies $\tilde d_{\rho,m} = (-1)^m d_{\rho,-m}$.
The model parameters in Eqs. (\ref{eqn:H-IBM}) and (\ref{eqn:Q-rho}) are denoted as $\epsilon_d$, $\kappa$, $\xi_1$, $\xi_2$, $\xi_3$, $\chi_\rho = \chi_\nu$ or $\chi_\pi$, $c_\nu^{(0)}$, $c_\nu^{(2)}$ and $c_\nu^{(4)}$.
They are fitted to reproduce the energy levels of the even-even core nucleus \cite{NUDAT,Kurpeta:2010zza,Kurpeta:2018zia}.

$H^{\rm F}_\rho$ is the odd-fermion Hamiltonian \cite{Iachello:2005aqa,Yoshida:2013jh},
\begin{equation}
	\label{eqn:ham-f}
	\begin{array}{rcl}
	H^{\rm F}_\rho = \displaystyle \sum_{j_\rho} \epsilon_{j_\rho} \hat n_{j_\rho} \mbox{ },
	\end{array}
\end{equation}
where $\hat n_{j_\rho}$ is a number operator and
\begin{equation}
	\label{eqn:epsilon-j}
	\epsilon_{j_\rho} = \sqrt{(E_{j_\rho} - \lambda_\rho)^2 + \Delta^2}
\end{equation}
is the quasi-particle energy of the odd particle, calculated in the BCS approximation \cite{Bardeen:1957mv,Alonso:1984rvl,Arias:1985,Alonso:1986,Arias:1985kjc}. Here, $\Delta = 12 /\sqrt A$ MeV is the pairing gap energy \cite{Bohr-Mottelson}, $\lambda_\rho$ the Fermi energy, and $E_{j_\rho}$ the proton/neutron single-particle energy for orbital $j$.

Finally, $V^{\rm BF}_\rho$ is the boson-fermion Hamiltonian, describing the interaction between the odd nucleon and the even-even nucleus. One has \cite{Iachello:2005aqa,Yoshida:2013jh}
\begin{widetext}
\begin{equation}
	\label{eqn:ham-bf}
	\begin{array}{rcl}	
	V^{\rm BF}_\rho & = & \displaystyle \sum_{i,j} \left\{ \Gamma_{ij} \left( \left[a^\dag_i\times\tilde a_j\right]^{(2)} 
	Q_{\rho'}^{\rm B} \right) + \left[ \Lambda^{j}_{ki} \Bigl( : \bigl[ [ d_{\rho}^{\dagger} \times \tilde{a}_{j} ]^{(k)}
	\times [a_{i}^{\dagger} \times s_{\rho}] \bigr]^{(2)} : \cdot \bigl[s_{\rho'}^{\dagger} \times \tilde{d}_{\rho'}\bigr]^{(2)} \Bigr)
	+ \text{ H.c.} \right] \right\} \\
	& + & A \displaystyle  \sum_{i} \hat{n}_{i} \hat{n}_{d_{\rho'}}
	\mbox{ },
	\end{array}
\end{equation}
\end{widetext}
where $\rho'\neq\rho$ indicates the other type of nucleon, i.e. $\rho' = \nu$ when $\rho = \pi$ and vice versa; $a^\dag_{i,j}$ are fermion creation operators.
The orbital dependence of the interaction strengths is parametrized according to \cite{Iachello:2005aqa,scholten,Iachello:1979zz,Yoshida:2013jh}
\begin{equation}
	\label{eq:gammaij}
	\Gamma_{i,j} = (u_{i}u_{j}-v_{i}v_{j}) \, Q_{i,j} \, \Gamma   \mbox{ }
\end{equation}	
and
\begin{equation}
	\label{eq:lambdaijk}
	\Lambda^{j}_{k,i} = -\beta_{k,i} \beta_{j,k} \left( \frac{10}{N_{\rho}(2j_{k}+1)} \right)^{1/2} \Lambda \mbox{ } ,
\end{equation}
where
\begin{equation}
	\label{eq:beta2}
	\beta_{i,j} = (u_{i}v_{j}+v_{i}u_{j}) \, Q_{i,j}  
\end{equation}
and
\begin{equation}
 	\label{eq:q2}
	\begin{array}{rcl}	
	Q_{i,j} & = & \langle l_{i},\tfrac{1}{2},j_{i} || Y^{(2)} || l_{j},\tfrac{1}{2},j_{j} \rangle \\
	& = & \frac{1+(-1)^{l_{i}+l_{j}}}{2} \sqrt{\frac{5(2j_{i}+1)}{4\pi}} \left( j_{i} \tfrac{1}{2} \; 2 \, 0 | j_{j} \tfrac{1}{2} \right)  		\mbox{ }.
	\end{array}
\end{equation}
In Eqs. (\ref{eqn:ham-bf}-\ref{eq:lambdaijk}), $A$, $\Gamma$ and $\Lambda$ are model parameters, which need to be fitted to reproduce experimental data. Occupation probabilities of the orbital $j$, $u_j$ and $v_j$, satisfy the relation $u_j^2 + v_j^2 = 1$.

A diagonalization of Eq. (\ref{eqn:H-IBFM}) in the boson-fermion space is unfeasible, because the dimension of the matrix Hamiltonian is expected to be extremely large.
Thus, the usual strategy is to carry out a pre-diagonalization of the core Hamiltonian, Eq. (\ref{eqn:H-IBM}), in the boson space.
After doing that, one can couple the single particle orbits of the odd nucleon to the lowest core-nucleus eigenstates; then, one can diagonalize the $H_{\rm F}^{\nu,\pi} + V_{\rm BF}^{\nu,\pi}$ interaction on the truncated basis and fit the model parameters of Eq. (\ref{eqn:ham-bf}) to the odd-$A$ nucleus energy levels.

\subsection{Even-even $^{116,118}$Pd core nuclei}
\label{Even-even 116,118Pd core nuclei}
The first step in the calculation of the odd-$A$ isotope energy levels is to calculate the properties of their even-even core nuclei.
In particular, we need an IBM-2 description of $^{118}$Pd, which is the core nucleus of both $^{117}$Pd and $^{117}$Rh, and of $^{116}$Pd, which is the core of $^{115}$Pd and $^{115}$Rh.
\begin{table}[htbp]
\centering
\begin{ruledtabular}
\begin{tabular}{ccccc}
Nucleus & $\epsilon_{\rm d}$ & $\kappa$  & $\chi_\pi$ & $\chi_\nu$ \\
\hline
$^{116}$Pd & 0.590 & $-0.150$  & $-0.300$ & 0.300 \\
$^{118}$Pd & 0.620 & $-0.175$  & $-0.400$ & 0.445 \\
\hline
Nucleus & $c_2$ & $c_4$ & $\xi_1$ & $\xi_2$ \\
\hline
$^{116}$Pd & $-0.100$ & 0.100 & 0.200 & 0.050 \\
$^{118}$Pd & $-0.100$ & 0.100 & 0.200 & 0.050 \\
\end{tabular}
\end{ruledtabular}
\caption{IBM-2 model parameters for the $^{116}$Pd and $^{118}$Pd cases from \cite[Tables 2 and 8]{Kim:1996aua}. All the values are in MeV, with the exception of those of $\chi_\pi$ and $\chi_\nu$ which are dimensionless. The IBM-2 parameters not shown here are set to zero.}
\label{tab:IBM-2-Model-parameters}
\end{table}
\begin{figure}[htbp]
\begin{center}
\includegraphics[width=7.5cm]{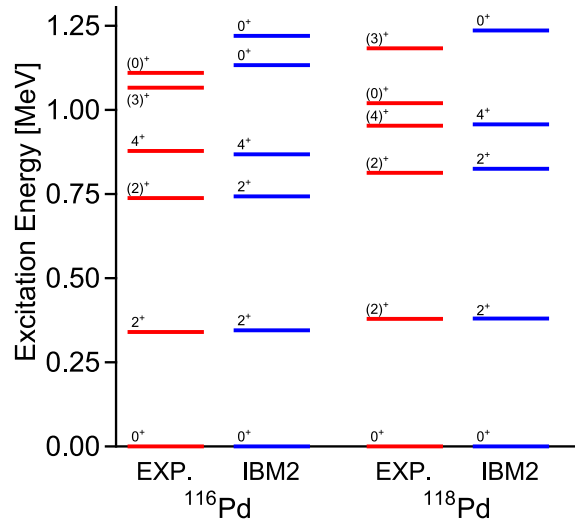}
\end{center}
\caption{The IBM-2 results for the $^{116}$Pd and $^{118}$Pd energy levels \cite{Kim:1996aua} are compared to the existing experimental data \cite{NUDAT,Kitao:1995iul,Blachot:2010jqe}.} 
\label{fig:IBM-2-spectra}
\end{figure} 

The even-even $^{(116,118)}$Pd nuclei were studied in the context of the IBM-2 in Ref. \cite{Kim:1996aua}, where the authors calculated excitation energies, electromagnetic transition strengths and electromagnetic moments in chain of  even-even Pd isotopes.
In the IBM-2, $^{116}$Pd, with $Z = 46$ and $N = 70$, is described in terms of 2 $\pi$-type and 6 $\nu$-type bosons, in both cases treated as holes; $^{118}$Pd, with $Z = 46$ and $N = 72$, is described in terms of 2 $\pi$-type and 5 $\nu$-type bosons, again treated as holes.
The IBM-2 model parameters for the $^{116}$Pd and $^{118}$Pd nuclei are reported in Table \ref{tab:IBM-2-Model-parameters}; see also \cite[Tables 2 and 8]{Kim:1996aua}. In Fig. \ref{fig:IBM-2-spectra} we compare the IBM-2 results of Ref. \cite{Kim:1996aua} to the experimental data \cite{NUDAT,Kitao:1995iul,Blachot:2010jqe}.

\begin{table}
\centering
\begin{ruledtabular}
\begin{tabular}{cccccc}
Nucleus & $1g_{7/2}$ & $2d_{5/2}$ & $2d_{3/2}$ & $3s_{1/2}$ & $1h_{11/2}$ \\
\hline
$^{115}$Pd & $0.883$ & $-0.015$ & $2.859$ & $2.146$ & $2.513$ \\
$^{117}$Pd & $0.858$ & $-0.008$ & $2.830$ & $2.169$ & $2.507$ \\
\end{tabular}
\end{ruledtabular}
\caption{Neutron single-particle energies (in MeV) of $^{115}$Pd and $^{117}$Pd isotopes used in the present IBFM-2 model calculations. The energies are extrapolated from the results of \cite[Table IV]{Yoshida:2002ju}.}
\label{tab:single-particle-energies-extrapolated}
\end{table}

\begin{figure*}
\begin{minipage}{18pc}
\includegraphics[width=18pc]{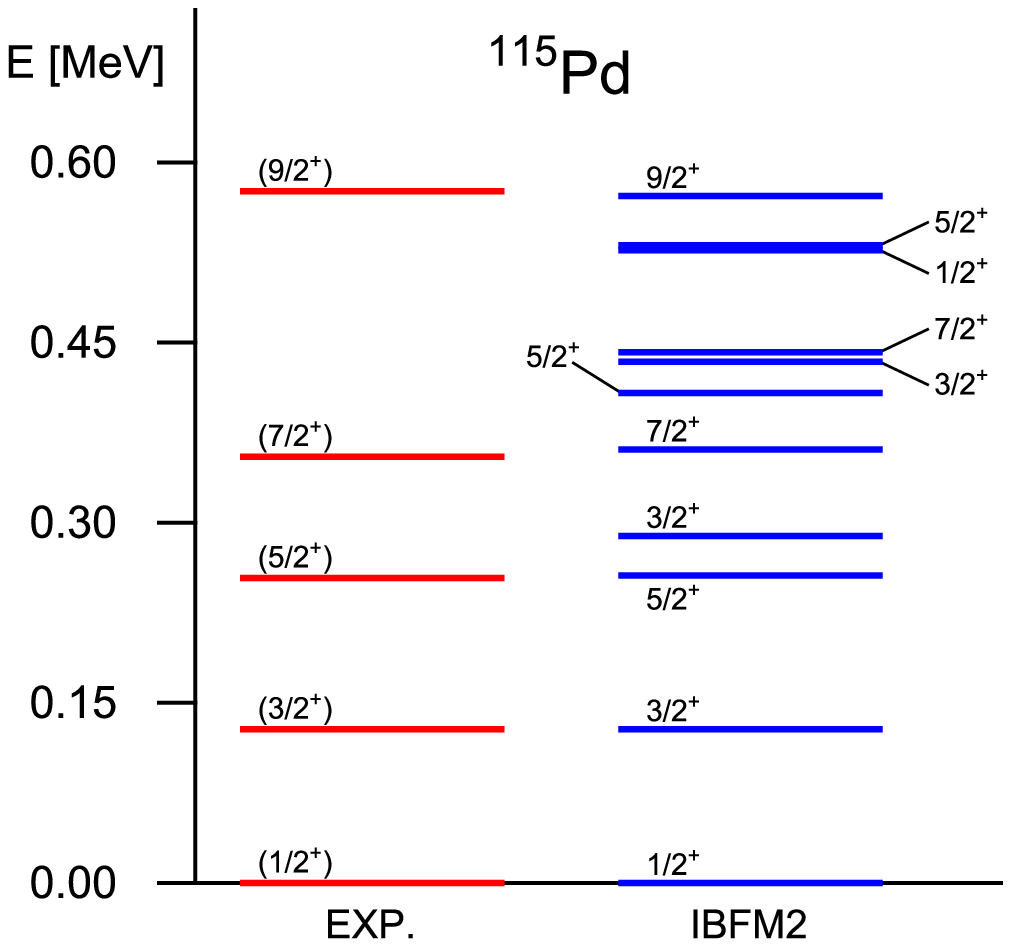}
\end{minipage}
\hspace{3pc}
\begin{minipage}{18pc}
\includegraphics[width=18pc]{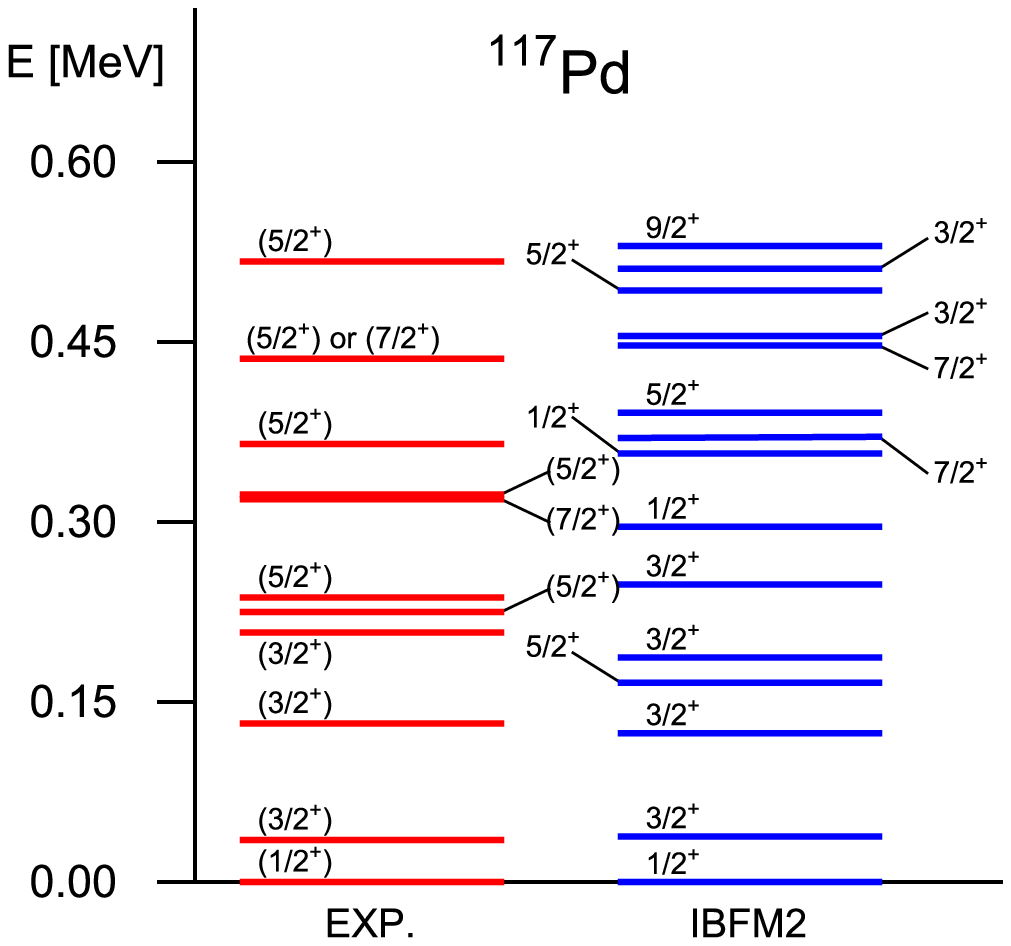}
\end{minipage}
\caption{Our IBFM-2 results for the positive-parity $^{115}$Pd (left) and $^{117}$Pd (right) energy levels are compared to the most recent experimental data \cite{Kurpeta:2010zza,Kurpeta:2018zia}.}
\label{fig:IBFM-2-115117Pd}
\end{figure*} 

\subsection{Spectra of the $^{(115,117)}$Pd isotopes in the IBFM-2}
\label{Spectra of the 115,117Pd isotopes in the IBFM-2}
The next step in the IBFM-2 procedure to evaluate the energy levels of the $^{(115,117)}$Pd isotopes is to calculate the neutron quasi-particle energies $\epsilon_{j_\nu}$ of Eq. (\ref{eqn:epsilon-j}).

The neutron quasi-particle energies are calculated by solving the BCS equations with the orbitals belonging to the 50-82 shell displayed in Table \ref{tab:single-particle-energies-extrapolated}. The BCS calculation requires unperturbed neutron single-particle energies (SPEs), $E_{j_\nu}$, as inputs.
The neutron SPEs of the $^{(115,117)}$Pd isotopes are extrapolated from the results of \cite[Table IV]{Yoshida:2002ju} for $^{(105-109)}$Pd. The outcome of the extrapolation is reported in Table \ref{tab:single-particle-energies-extrapolated}.
The results of \cite[Table IV]{Yoshida:2002ju} were extracted from \cite{Reehal:1970yv}, except for the energies of the $g_{7/2}$ orbit which were slightly lowered. 

The last step of the IBFM-2 procedure is to couple the odd-nucleon to the even-even core nucleus, see Sec. \ref{Even-even 116,118Pd core nuclei}, and then to diagonalize the IBFM-2 Hamiltonian.
By fitting the IBFM-2 model parameters to reproduce the most recent experimental data for the $^{(115,117)}$Pd spectra \cite{Kurpeta:2010zza,Kurpeta:2018zia}, one obtains both the IBFM-2 predictions for the $^{(115,117)}$Pd energy levels and their wave functions.
Our results, calculated via the IBFM-2 Hamiltonian of Eq. (\ref{eqn:H-IBFM}) and the model parameters of Table \ref{tab:IBFM-2-Pd-parameters}, are shown in Figs. \ref{fig:IBFM-2-115117Pd}.
\begin{table}
\centering
\begin{ruledtabular}
\begin{tabular}{cccc}
Nucleus & $\Gamma$ [MeV] & $\Lambda$ [MeV] & $A$ [MeV] \\
\hline
$^{115}$Pd & 0.77 & 0.21 & $-0.28$ \\
$^{117}$Pd & 0.16 & 0.64 & $-0.36$ \\
\end{tabular}
\end{ruledtabular}
\caption{IBMF2 model parameters for the $^{(115,117)}$Pd isotopes, obtained by fitting the IBFM-2 model parameters to the most recent experimental data \cite{Kurpeta:2010zza,Kurpeta:2018zia}.}
\label{tab:IBFM-2-Pd-parameters}
\end{table}

\subsection{Spectra of the $^{(115,117)}$Rh isotopes in the IBFM-2}
The procedure to compute the energy levels of the $^{(115,117)}$Rh isotopes in the IBFM-2 is substantially the same as that of Sec. \ref{Spectra of the 115,117Pd isotopes in the IBFM-2}.

The single particle energies for the $^{115}$Rh and $^{117}$Rh isotopes are obtained by solving the Schr\"odinger equation for the Woods-Saxon (WS) potential; this is the sum of a spin-independent central, a spin-orbit, and a Coulomb part \cite{Brown}. We make use of a typical set of WS parameters. In particular, the strength parameters are: $V_{0} = - 53$ MeV, $V_{1}= - 30$ MeV and $V_{\rm so} = 22$ MeV; for the geometry, we have: $r_{0} = r_{\rm so} = 1.3$ fm, and $a_{0} = a_{\rm s0} = 0.7$; the radius of the Coulomb term is $r_{\rm c} = 1.20$ fm.
The resulting proton SPEs are reported in Table \ref{tab:single-particle-energies2}.
\begin{table}[htbp]
\centering
\begin{ruledtabular}
\begin{tabular}{cccccccc}
Nucleus & $2p_{1/2}$ & $2p_{3/2}$ & $1f_{5/2}$ & $1g_{9/2}$ & $2d_{5/2}$ & $1g_{7/2}$ \\
\hline
$^{115}$Rh & 6.917 & 8.554 & 9.029 & 6.481 & 1.075 & 0.000  \\
$^{117}$Rh & 6.818 & 8.429 & 8.975 & 6.367 & 0.946 & 0.000  \\
\end{tabular}
\end{ruledtabular}
\caption{Proton single-particle energies of $^{115}$Rh and $^{117}$Rh isotopes used in the present IBFM-2 model calculations. The results are obtained in a WS potential calculation, with the values of the model parameters reported in the text.}
\label{tab:single-particle-energies2}
\end{table}

\begin{figure*}[htbp]
\begin{minipage}{18pc}
\includegraphics[width=18pc]{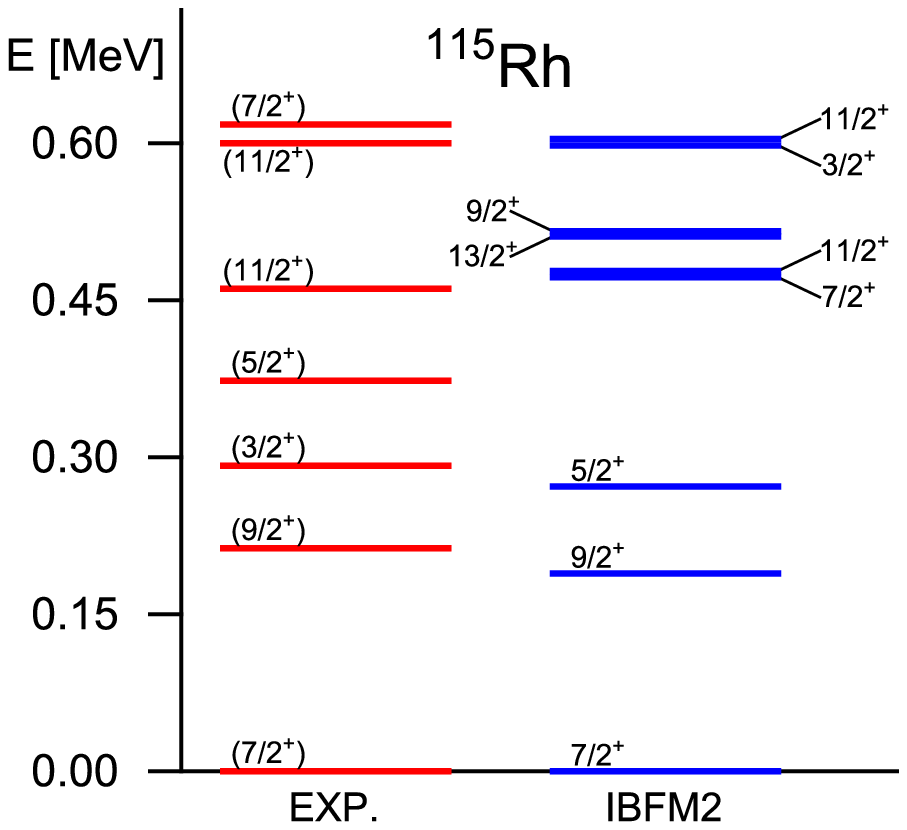}
\end{minipage}
\hspace{3pc}
\begin{minipage}{18pc}
\includegraphics[width=18pc]{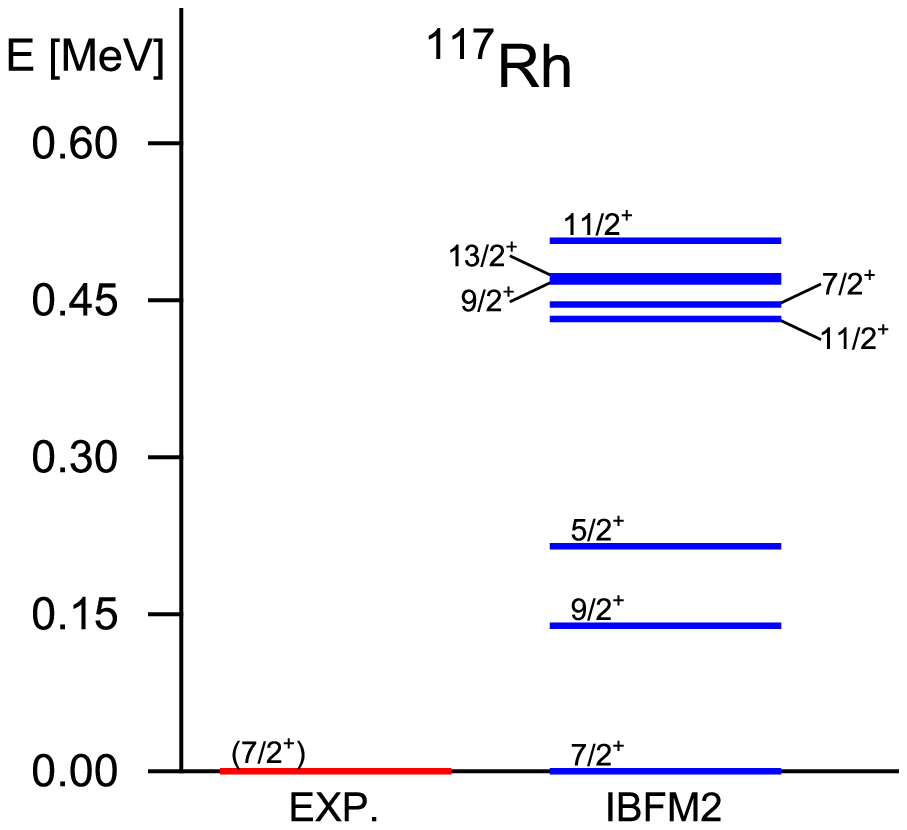}
\end{minipage}
\caption{Our IBFM2 results for the positive-parity $^{115}$Rh (left) and $^{117}$Rh (right) energy levels are compared to the existing experimental data \cite{NUDAT}.}
\label{fig:IBFM2-115117Rh}
\end{figure*} 

Finally, we can couple the odd-proton to the even-even core nucleus and diagonalize the IBFM-2 Hamiltonian.
We choose to use the same parameters both in the $^{115}$Rh and $^{117}$Rh cases because of the lack of experimental data for these isotopes. Our calculation reproduces the tentative $J^P = \frac{7}{2}^+$ quantum number assignment to the ground-states of both isotopes.
The resulting energy levels are shown In Figs. \ref{fig:IBFM2-115117Rh} and are also used in the calculations of Sec. \ref{beta decays}.

\begin{table}[htbp]
\centering
\begin{ruledtabular}
\begin{tabular}{cccc}
Nucleus & $\Gamma$ [MeV] & $\Lambda$ [MeV] & $A$ [MeV] \\
\hline
$^{115}$Rh & $-0.50$ & 0.75 & $-2.00$ \\
$^{117}$Rh & $-0.50$ & 0.75 & $-2.00$ \\
\end{tabular}
\end{ruledtabular}
\caption{As Table \ref{tab:IBFM-2-Rh-parameters}, but for the $^{(115,117)}$Rh isotopes.}
\label{tab:IBFM-2-Rh-parameters}
\end{table}

\section{$M1$ transitions of $^{(115,117)}$Pd isotopes in the IBFM-2 and their ground-state quantum number assignments}
\label{sect3}
By solving the eigenvalue problem of Eq. (\ref{eqn:H-IBFM}), one gets both the energy levels and the wave functions of the nuclei of interest.
The numerical wave functions can then be used to study the electromagnetic properties of those nuclei in the IBFM-2.

By making use of the spectator approximation, the one-body $M1$ electromagnetic transition operator can be written as \cite{Yoshida:2002ju,Iachello:2005aqa}
\begin{equation}
	\label{eq:m1}
	\begin{array}{rcl}
	T^\mathrm{(M1)} & = & \displaystyle \sum_{\rho=\nu,\pi} g_\rho^{\rm B} [d_\rho^\dag \times \tilde d_\rho 
	+ \mbox{h.c.}]^{(1)} \\
	&+& \sqrt{\frac{3}{4\pi}} \displaystyle \sum_{i,j} e_{ij}^{(1)} [a_i^\dag \tilde a_j]^{(1)}  \mbox{ },
	\end{array}
\end{equation}
where $g_\rho^\mathrm{B}$  is the boson $g$-factor for the neutron/proton boson. 
\begin{equation}
	\begin{array}{l}
	\label{eqn:eij1}	
	e_{ij}^{(1)}= - \frac{1}{\sqrt 3} (u_i u_j + v_i v_j) \left\langle l_i \frac{1}{2}j_i \middle\| g_\ell {\bm \ell} 
	+ g_s {\bf s} \middle\| l_j \frac{1}{2}j_j \right\rangle  
	\end{array}
\end{equation}
is the fermion single-particle matrix element, where $g_\ell$ and $g_s$ are the single-particle $g$-factors.
The values of the boson $g$-factors of Eq. (\ref{eq:m1}) are extracted from Ref. \cite{Kim:1996aua}, where the values $g_\nu^{\rm B} = - 0.1$ and $g_\pi^{\rm B} = 1.25$ are used.
For the single-particle $g$-factors the bare values $g_\ell=0$ and $g_s=-3.826$ are employed. See also the discussion on the effective charges in \cite[Sec. IIA]{Nomura:2017efu}.

In what follows, we use our IBFM-2 predictions for $M1$ transitions to discuss the $J^P$ quantum number assignments for the $^{(115,117)}$Pd ground-states; the experimental data can be found in Refs. \cite{Aysto:1988axd,Penttila-PhD,Kurpeta:2010zza,Kurpeta:2018zia,Penttila:1991hd,Urban:2004,Fong:2005ey,Lalkovski:2013pba}.
This critical analysis can be useful due to the conflicting nature of the previous experimental results.
Specifically, Ref. \cite{Penttila:1991hd} suggested a $J^P = \frac{5}{2}^+$ assignment for the $^{117}$Pd ground-state (plus a $J^P = \frac{7}{2}^+$ assignment for that of $^{117}$Rh); Ref. \cite{Urban:2004} indicated $J^P = \frac{3}{2}^+$ assignments for the ground-states of both $^{115}$Pd and $^{117}$Pd, while Ref. \cite{Fong:2005ey} gave $J^P = \frac{1}{2}^+$ assignments for $^{115}$Pd and $J^P = \frac{3}{2}^+$ for $^{117}$Pd and most recently Refs. \cite{Kurpeta:2010zza,Kurpeta:2018zia} reported $J^P = \frac{1}{2}^+$ ground-states for both $^{115}$Pd and $^{117}$Pd  followed by one/two $J^P = \frac{3}{2}^+$ excitations.
Features of the latest experiment are reproduced by our theoretical results; see Fig. \ref{fig:IBFM-2-115117Pd}.
\begin{figure}[htbp]
\begin{center}
\includegraphics[width=7.5cm]{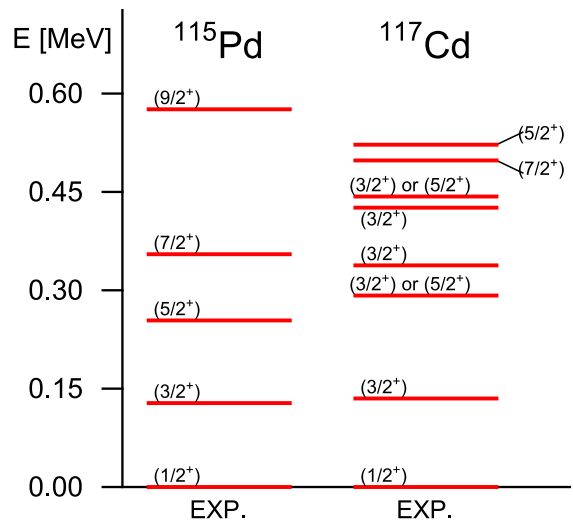}
\end{center}
\caption{Comparison between the experimental energy levels of $^{115}$Pd \cite{Kurpeta:2010zza} and $^{117}$Cd \cite{ENSDF} isotones.} 
\label{fig:115-117-comparison}
\end{figure} 

Because of the lack of experimental data for the $M1$ transitions of $^{(115,117)}$Pd isotopes, we use that of $^{117}$Cd as a reference \cite{ENSDF} to get a hint of the situation.\footnote{We refer to $^{117}$Cd experimental results because Ref. \cite{Kurpeta:2018zia} only provides the relative intensities of $^{117}$Pd electromagnetic transitions; it does not give the absolute intensities of either $E2$ or $M1$ transitions of $^{117}$Pd.}
The spectrum of $^{117}$Cd \cite{ENSDF} is indeed very similar to that of $^{115}$Pd \cite{Kurpeta:2010zza}, because the $^{115}$Pd and $^{117}$Cd nuclei are characterized by the same number of neutrons; it also shares some features with the spectrum of $^{117}$Pd. See Fig. \ref{fig:115-117-comparison}.

Ref.~\cite{ENSDF} gives experimental lower limit  for the $\frac{3}{2}^+_1 \rightarrow \frac{1}{2}^+_1$ $M1$ transition in $^{117}$Cd nucleus
\begin{equation}
	B\left(M1; \frac{3}{2}^+_1 \rightarrow \frac{1}{2}^+_1\right) > 0.0074  \text{ W.u}.
\end{equation}	
In the IBFM-2, we get
\begin{equation}
	B\left(M1; \frac{3}{2}^+_1 \rightarrow \frac{1}{2}^+_1\right) = 0.009  \text{ W.u}\mbox{ }
\end{equation}
and 
\begin{equation}
	\frac{B\left(M1; \frac{3}{2}^+_2 \rightarrow \frac{3}{2}^+_1\right)}{B\left(M1; \frac{3}{2}^+_2 \rightarrow \frac{1}{2}^+_1\right)} = 0.17  \mbox{ }
\end{equation}
in the case of $^{117}$Pd, Ref. \cite{Kurpeta:2018zia} reports relative intensities for $\frac{3}{2}^+_2 \rightarrow \frac{1}{2}^+_1$ and $\frac{3}{2}^+_2 \rightarrow \frac{3}{2}^+_1$ lines, those being $100(1)$ and $34.9(1.6)$,   respectively.

In case of $^{115}$Pd, the corresponding IBFM-2 values are
\begin{equation}
	B\left(M1; \frac{3}{2}^+_1 \rightarrow \frac{1}{2}^+_1\right) = 0.016  \text{ W.u}\mbox{ }
\end{equation}
and 
\begin{equation}
	\frac{B\left(M1; \frac{3}{2}^+_2 \rightarrow \frac{3}{2}^+_1\right)}{B\left(M1; \frac{3}{2}^+_2 \rightarrow \frac{1}{2}^+_1\right)} = 0.55  \mbox{ }.
\end{equation}

Our findings are compatible with the assignments of Refs. \cite{Kurpeta:2010zza,Kurpeta:2018zia} for the ordering of the lowest-lying $^{115}$Pd and $^{117}$Pd levels.

\section{$\beta$-decay rates of $^{(115,117)}$Rh into $^{(115,117)}$Pd isotopes in the IBFM-2}
\label{beta decays}
The last step of our study is the calculation of the $^{(115,117)}$Rh $\rightarrow$ $^{(115,117)}$Pd $\beta$-decay rates.
At first, we briefly discuss how to compute the $\beta$-decay rates in the IBFM-2. Further details can be found elsewhere \cite{Dellagiacoma:1989fii,Dellagiacoma:Thesis}.

The $\beta$-decay half-lives can be calculated as \cite{Suhonen:Nuclear}
\begin{equation}
	\label{eqn:t1/2}
	t_{1/2} = \frac{\kappa}{f_0 \left( \langle M_{\rm F} \rangle^2 
	+ \left( \frac{g_A}{g_V}\right)^2 \langle M_{\rm GT} \rangle^2 \right)}  \mbox{ };
\end{equation}	
here, $f_0$ is a leptonic phase-space factor; $\kappa = \frac{2\pi^2\hbar^7\mbox{ }{\rm ln }2}{m_e^5c^4G_F^2} = 6163$ s and $\frac{g_A}{g_V} = -1.2756\pm0.0030$ \cite{Mendenhall:2012tz} are constants; $\langle M_{\rm F} \rangle$ and $\langle M_{\rm GT} \rangle$ are the matrix elements of Fermi, $T_{\rm F}$, and Gamow-Teller, $T_{\rm GT}$, operators between the wave functions of the parent and daughter nuclei. 
\begin{table}
\centering
\begin{ruledtabular}
\begin{tabular}{cccc}
$\beta$ transition & $(J_{\rm i})_{n_{\rm i}} \rightarrow (J_{\rm f})_{n_{\rm f}}$ & log$_{10}ft^{\rm exp}$ & log$_{10}ft^{\rm th}$ \\
\hline
$^{115}$Rh $\rightarrow$ $^{115}$Pd & $(\frac{7}{2})_1 \rightarrow (\frac{5}{2})_1$ & -- & 5.90 \\
                                                             & $(\frac{7}{2})_1 \rightarrow (\frac{5}{2})_2$ & -- & 5.83 \\
                                                             & $(\frac{7}{2})_1 \rightarrow (\frac{5}{2})_3$ & -- & 6.85 \\
                                                             & $(\frac{7}{2})_1 \rightarrow (\frac{9}{2})_1$ & -- & 7.23 \\
                                                             & $(\frac{7}{2})_1 \rightarrow (\frac{9}{2})_2$ & -- & 7.78 \\
                                                             & $(\frac{7}{2})_1 \rightarrow (\frac{9}{2})_3$ & -- & 7.37 \\
\hline                                                                                                                          
$^{117}$Rh $\rightarrow$ $^{117}$Pd & $(\frac{7}{2})_1 \rightarrow (\frac{5}{2})_1$ & 5.7 & 6.78 \\
                                                             & $(\frac{7}{2})_1 \rightarrow (\frac{5}{2})_2$ & 6.3 & 6.55 \\
                                                             & $(\frac{7}{2})_1 \rightarrow (\frac{5}{2})_3$ & 5.1 & 5.33 \\                                                             
                                                             & $(\frac{7}{2})_1 \rightarrow (\frac{7}{2})_1$ & 5.8 & 6.68 \\
                                                             & $(\frac{7}{2})_1 \rightarrow (\frac{7}{2})_2$ & 6.0$^\dag$ & 6.95 \\
                                                             & $(\frac{7}{2})_1 \rightarrow (\frac{7}{2})_3$ & --    & 7.44 \\                                                                                                                                                                                                                                                    
                                                             & $(\frac{7}{2})_1 \rightarrow (\frac{9}{2})_1$ & $6.2^\ddag$ & 8.28 \\
                                                             & $(\frac{7}{2})_1 \rightarrow (\frac{9}{2})_2$ & -- & 7.70 \\
                                                             & $(\frac{7}{2})_1 \rightarrow (\frac{9}{2})_3$ & -- & 6.85 \\                                                             
\end{tabular}
\end{ruledtabular}
\caption{log$_{10}ft$ values for Rh to Pd decays. The experimental results are extracted from Ref. \cite{Kurpeta:2018zia}. The notation $(J_{\rm i,f})_{n_{\rm i,f}}$ indicates the spin and radial quantum number of the initial/final nucleus. The experimental result marked as $\dag$ may correspond to the transition either to a $\frac{5}{2}^+$ or $\frac{7}{2}^+$ state, that marked as $\ddag$ may correspond to the transition either to a $\frac{7}{2}^+$ or $\frac{9}{2}^+$ state.}
\label{tab:LogFt}
\end{table}
If we consider, for example, the $\beta^-$ case, these operators are defined as \cite{Dellagiacoma:1989fii,Dellagiacoma:Thesis,Mardones:2016wgy}
\begin{equation}
	\label{eqn:TF}
	T_{\rm F} = - \displaystyle \sum_i {\hat j}_i \left[ A_{\pi,i} c^\dag_{\pi,i} \times A_{\nu,i} \tilde c_{\nu,i}\right]^{(0)}
\end{equation}
and
\begin{equation}
	\label{eqn:TGT}
	T_{\rm GT} = - \displaystyle \sum_{i,j} \eta_{ij} \left[ A_{\pi,i} c^\dag_{\pi,i} \times A_{\nu,j} \tilde c_{\nu,j}\right]^{(1)}_\mu
	\mbox{ },
\end{equation}
where the index $i$ ($j$) denotes a particular shell, characterized by the standard single-particle level quantum numbers $n_i, \ell_i, \frac{1}{2}, j_i, m_i$, and $\hat j_i = \sqrt{2 j_i +1}$.
The quantity $\eta_{ij}$ is defined as \cite{Dellagiacoma:1989fii,Dellagiacoma:Thesis}
\begin{equation}
	\eta_{ij} = \sqrt 2 (-1)^{\ell_i+j_i+\frac{1}{2}} \hat j_i \hat j_j 
	\left\{ \begin{array}{rcl} \frac{1}{2} & \frac{1}{2} & 1 \\ j_j & j_i & \ell_i \end{array}\right\}  \delta_{\ell_i,\ell_j}  \mbox{ }.
\end{equation}
The operators $c^\dag_{\rho,i}$ and $\tilde c_{\rho,i}$ (with $\rho = \nu, \pi$) in Eqs. (\ref{eqn:TF}) and (\ref{eqn:TGT}) are the so-called transfer operators. They can create/destroy a nucleon in the parent and daughter nuclei.
The values of the coefficients $A_{\rho,i}$ are calculated by means of the OAI method \cite{Otsuka:1978zz} and depend on the specific normalization one takes into account. 
Here, we use the conventions for $A_{\rho,i}$s reported in Refs. \cite{Mardones:2016wgy,Barea:2014lza}, which are based on the procedure for diagonalizing the Surface Delta Interaction (SDI) of Ref. \cite{Pittel:1982} and the use of the commutator method of Refs. \cite{Frank:1982zz,Lipas:1990rs}.
The OAI method is based on the Generalized Seniority (GS) scheme in the Shell Model (SM).
In this scheme, the SM space is truncated to the $SD$ pair space.
One has \cite{Otsuka:1978zz}
\begin{equation}
	\label{eqn:SandD}
	S^\dag = \displaystyle \sum_{i=1}^k \frac{\alpha_i \hat j_i A_{ii}^{(00)}}{2}  \mbox{ },
	\mbox{ } \mbox{ } D^\dag_\mu = \displaystyle \sum_{i,i'=1, i \le i'}^k \frac{\beta_{ii'} A_{ii'}^{(2\mu)}}{\sqrt{1+\delta_{ii'}}}
	\mbox{ },
\end{equation}	
where $A_{ii'}^{(2\mu)} = (c_i^\dag \times c_{i'}^\dag)^{(J)}_M$, $c_{i,i'}^\dag$ being one-nucleon transfer operators, which can be written in terms of fermion and boson creation/annihilation operators \cite[Eq. (6)]{Barea:2014lza}; $\alpha$s and $\beta$s are the pair structure coefficients of the $S$ and $D$ pairs, respectively.
$\alpha$s and $\beta$s are obtained by diagonalizing the Surface Delta Interaction (SDI) \cite{Pittel:1982} and are normalized according to
\begin{equation}
	\displaystyle \sum_j \alpha_j^2 \Omega_j = \displaystyle \sum_j \Omega_j  \mbox{  } \mbox{ and }
	\sum_{i,i'=1, i \le i'}^k \beta_{ii'}^2 = 1 \mbox{ }.
\end{equation}	
While the overall sign of these coefficients is not relevant, the relative sign is important.
To that purpose, one can use the approximate relation
\begin{equation}
	\beta_{i,i'} = \frac{\alpha_i \alpha_{i'}}{\sqrt 5 \Omega (1 + \delta_{ii'})} 
	\left\langle l_i \frac{1}{2}j_i \middle\| r^2 Y^{(2)} \middle\| l_{i'} \frac{1}{2}j_{i'} \right\rangle \mbox{ },
\end{equation}
with $\Omega = \displaystyle \sum_j \Omega_j$ to extract the $\beta_{i,i'}$ directly from the $a_i$ coefficients.	
Finally, by making use of the commutator method introduced by Frank and Van Isacker \cite{Frank:1982zz} (see also Ref. \cite{Lipas:1990rs}), one can extract the exact value of the occupation probability $v_j^2$ \cite[Eq. (5)]{Barea:2014lza}. 
One can also calculate the matrix elements of the $(a^\dag_i \times \tilde a_{i'})^{(K)}$ operators, which are important to extract the expressions of the mapping of Eq. (\ref{eqn:SandD}) and the coefficients of the one-nucleon transfer operator.
As discussed in Ref. \cite{Matus:2017eni}, the OAI + SDI mapping is particular effective in spherical and vibrational regions, where one considers low GS states.

Instead of the quantities of Eq. (\ref{eqn:t1/2}), one usually calculates the so-called $ft$ values, which are defined as the product of $f_0$ and $t_{1/2}$; one has:
\begin{equation}
	\label{eqn:ft}
	ft = \frac{\kappa}{\left( \langle M_{\rm F} \rangle^2 
	+ \left( \frac{g_A}{g_V}\right)^2 \langle M_{\rm GT} \rangle^2 \right)}  \mbox{ }.
\end{equation}
The $ft$ values depend exclusively on the nuclear structure, i.e. the nuclear matrix elements $\langle M_{\rm F} \rangle$ and $\langle M_{\rm GT} \rangle$.

Finally, the results of our microscopic IBFM-2 calculation of the $ft$ values (or better their log$_{10}$ logarithms) are reported in Table \ref{tab:LogFt}. Our results are compared with the existing experimental data from Ref. \cite{Kurpeta:2018zia}.
It is worth to note that the experimental results for $^{117}$Rh $\rightarrow$ $^{117}$Pd $\beta$-transitions, or at least their general trend, are reasonably well reproduced by our IBFM-2 findings.

The comparison between our IBFM-2 predictions of energy levels, $M1$ electromagnetic transitions, and  ($\beta$-decays), with experimental data \cite{Aysto:1988axd,Penttila-PhD,Kurpeta:2010zza,Kurpeta:2018zia,Penttila:1991hd,Urban:2004,Fong:2005ey,Lalkovski:2013pba} seem to favor the $J^P = \frac{1}{2}^+$ quantum number assignments of Refs. \cite{Fong:2005ey,Kurpeta:2010zza,Kurpeta:2018zia} for the ground-states of $^{(115,117)}$Pd.

\section{Conclusions}
\label{conclusions}
We have studied the properties of the $^{(115,117)}$Rh and $^{(115,117)}$Pd isotopes, including their spectra and electromagnetic transitions, and calculated the Rh $\rightarrow$ Pd beta decays in the microscopic IBFM-2 formalism \cite{Iachello:2005aqa,Arima:1977vie,Otsuka:1978zza,Otsuka:1978zz}.
These type of investigations are important to get insight into the structure of heavy nuclei and test the available nuclear structure models. 
The study of $\beta$, double-$\beta$ and neutrinoless double-$\beta$ decays may also provide valuable informations on the properties of neutrinos and the possible emergence of beyond the Standard Model effects in weak interactions \cite{Giunti:2003qt}.

Our theoretical results are in good agreement with the latest experimental data \cite{Fong:2005ey,Kurpeta:2010zza,Kurpeta:2018zia} for the properties of the $^{(115,117)}$Rh and $^{(115,117)}$Pd isotopes and support the $J^P = \frac{1}{2}^+$ quantum number assignments of Refs. \cite{Fong:2005ey,Kurpeta:2010zza,Kurpeta:2018zia} for the ground-states of $^{(115,117)}$Pd.

\begin{acknowledgments}
The authors acknowledge useful discussions with Jos\'e Barea, Universidad de Concepci\'on, Chile.
This work was supported by the Academy of Finland, Grant Nos. 314733, 320062, and INFN, Italy.
\end{acknowledgments}

\end{document}